\documentclass[runningheads]{llncs}
\usepackage[T1]{fontenc}
\usepackage{graphicx}
\usepackage{amsmath,amssymb}
\usepackage{booktabs}
\usepackage{multirow}
\usepackage{url}
\usepackage{hyperref}
\usepackage[table]{xcolor}
\usepackage{xspace}
\newcommand{\AnyTwoReg}{Any\textsuperscript{2}Reg\xspace}

\newcommand{\mstd}[2]{#1$\pm$#2}

\newcommand{\bestcell}[1]{\cellcolor{gray!25}\textbf{#1}}
\newcommand{\secondcell}[1]{\cellcolor{gray!12}#1}
\newcommand{\failrun}{\textbf{-}}
\newcommand{\bestlegend}{\begingroup\setlength{\fboxsep}{0.8pt}\colorbox{gray!25}{\textbf{Best}}\endgroup}
\newcommand{\secondlegend}{\begingroup\setlength{\fboxsep}{0.8pt}\colorbox{gray!12}{Second}\endgroup}
\begin{document}

\title{Set-Based Groupwise Registration for Variable-Length, Variable-Contrast Cardiac MRI}
\titlerunning{\AnyTwoReg for Variable-Length and Contrast-Diverse Cardiac MRI}

\author{Yi Zhang\inst{1} \and Yidong Zhao\inst{1} \and Tijmen Toxopeus\inst{1} \and Ma\v{s}a Bo\v{z}i\'c-Iven\inst{1} \and Sebastian Weing\"artner\inst{1} \and Qian Tao\inst{1}}
\authorrunning{Y. Zhang et al.}
\institute{Department of Imaging Physics, Delft University of Technology, The Netherlands}

\maketitle

\begin{abstract}
Quantitative cardiac magnetic resonance imaging (MRI) enables non-invasive myocardial tissue characterization but relies on robust motion correction within these variable-length, variable-contrast image sequences. Groupwise registration, which simultaneously aligns all images, has shown greater robustness than pairwise registration for motion correction. However, current deep-learning-based groupwise registration methods cannot generalize across MRI sequences: the architecture typically encodes input data as a fixed-length channel stack, which rigidly couples network design to protocol-specific sequence length, input ordering, and contrast dynamics. At inference time, any change in imaging protocols will render the network unusable. In this work, we introduce \emph{\AnyTwoReg}, a new set-based groupwise registration framework that takes a quantitative MRI sequence as an unordered set. This set formulation fundamentally decouples network design from sequence length and input ordering. By utilizing a shared encoder and correlation-guided feature aggregation, \emph{\AnyTwoReg} constructs a permutation-invariant canonical reference for registration, and learns a permutation-equivariant mapping from images to deformation fields. Additionally, we extract contrast-insensitive image features from an existing foundation model to handle extreme contrast variations. Trained exclusively on a single public $T_1$ mapping dataset (STONE, sequence length $L=11$), \AnyTwoReg generalizes to two unseen quantitative MRI datasets (MOLLI, ASL) with variable lengths ($L \in [11, 60]$) and different contrast dynamics. It achieves strong cross-protocol generalization in a zero-shot manner, and consistently improves downstream quantitative mapping quality. Notably, while designed for quantitative MRI sequences, our framework is directly applicable to Cine MRI sequences for inter-cardiac-phase registration.
\end{abstract}

\section{Introduction}

Cardiac magnetic resonance imaging (MRI) is essential for diagnosing and monitoring cardiovascular diseases. Modern examinations typically include quantitative mapping, such as the widely used $T_1$ mapping~\cite{campello2021multi,bernard2018deep,messroghli2004modified,bovzic2024improved}, for non-invasive myocardial tissue assessment. Quantitative sequences estimate parameters by fitting signal models voxel-wise, assuming strict anatomical alignment across images. However, cardiac and respiratory motion cause misalignment, introducing spatial blurring, biased parameter estimates, and increased map uncertainty~\cite{kellman2013t1,kellman2014t1}. Motion correction is a prerequisite for reliable quantitative cardiac MRI~\cite{tao2018robust,xue2012motion}.

Quantitative cardiac MRI sequences are challenging to register. Their imaging protocols have variable sequence lengths $L$, specific sampling orders, and different contrast evolution governed by different physical processes~\cite {makela2002review,haaf2016cardiac}. Inter-frame differences reflect an entangled combination of true motion and physiological signal evolution. This poses significant challenges for traditional pairwise image registration, which requires a predefined reference image and fails for poor-contrast images~\cite{arava2021deep,li2022motion}. Therefore, groupwise registration is methodologically preferred, which aligns all images simultaneously to an implicit reference~\cite{huizinga2016pca,zhang2021groupregnet,zhang2024deep}. Classical groupwise registration performs optimization per sequence~\cite{huizinga2016pca}, and is computationally expensive. Importantly, the optimization scales poorly to sequence length and can take minutes to hours for long sequences. 

\begin{figure*}[b]
    \centering
    \includegraphics[width=\textwidth]{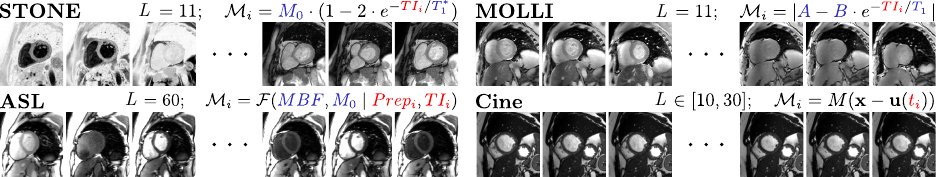}
    \caption{Cardiac MRI sequences exhibit high heterogeneity across imaging protocols with different mechanisms ($\mathcal{M}_i$) and sequence lengths ($L$). Analytical models are expressed by external parameters ({\color{red}{red}}; e.g., inversion times $TI_i$, preparation $Prep_i$, or phases $t_i$) and internal properties ({\color[RGB]{49,49,190}{dark blue}}; e.g., $T_1, T_1^*$, $MBF$, $M_0$, $A$, or $B$).}
    \label{data}
\end{figure*}

Deep learning provides an amortized alternative to per-sequence optimization, significantly accelerating inference~\cite{voxelmorph}. Recent works leverage deep learning to speed up groupwise registration for practical use~\cite{voxelmorph,martin2020groupwise,zhang2021groupregnet,li2022motion}. However, existing works predominantly use U-Net-style backbones~\cite{ronneberger2015u}, which hardcode the input image sequence as a fixed-length, ordered channel stack~\cite{li2023contrast,zhang2024deep,hanania2023pcmc,hanania2025mbss}. Such hard-coding of sequence length and order imposes a rigid prior, precluding generalization to other quantitative sequences with different lengths or contrasts. Though some recent works have explored variable-size group modelling for inter-subject brain MRI atlas construction~\cite{he2025instantgroup,abulnaga2025multimorph}, they primarily focus on homogeneous image contrasts and cannot address the dramatic contrast changes in cardiac MRI (e.g. contrast nulling, flipping, etc. as shown in Fig. \ref{data}). 

In principle, a quantitative MRI acquisition can be viewed as an observation set, each element being an image, sampled from a physical process (e.g., $T_1$ relaxation). Quantitative mapping operates on the set, where standard fitting (least-squares minimization) is invariant to permutations in the set. Ideally, motion correction models should also respect this set symmetry, but most groupwise registration methods use fixed-channel models~\cite{li2023contrast,zhang2024deep,hanania2023pcmc,hanania2025mbss}, where the sequence length and frame order critically determine learning, contradicting set symmetry. 

Motivated by this critical gap, we propose \AnyTwoReg, a set-based groupwise registration framework that treats cardiac MRI sequences as unordered sets. Instead of fixed-length channel stacking, \AnyTwoReg uses a shared encoder to extract features per image. It uses correlation-guided aggregation to construct a canonical reference, which is strictly permutation-invariant. \AnyTwoReg enables a new way of groupwise registration, generalizable to variable-length, variable contrast cardiac MRI sequences. Our contributions are threefold:
\begin{itemize}
    \item We introduce \AnyTwoReg, a set-based groupwise registration architecture that models quantitative MRI sequences as unordered sets. \AnyTwoReg fundamentally decouples network design from sequence length $L$ and acquisition order, enabling highly efficient and scalable registration of variable-length sequences with near-linear time complexity \textit{w.r.t.} $L$ in practice.
  \item We propose a correlation-guided aggregation module to construct a canonical reference, and an optional module for contrast-insensitive feature fusion. This makes \AnyTwoReg{} applicable to variable-contrast sequences. 
  \item We demonstrate that \AnyTwoReg, trained exclusively on a public $T_1$ database, achieves strong zero-shot generalization on three unseen cardiac MRI datasets with varying sequence lengths and image contrasts. It consistently outperforms state-of-the-art baselines in registration and downstream mapping. 
\end{itemize}
\section{Method}
\subsection{Problem formulation and set symmetry for registration}
Let $\Omega \subset \mathbb{R}^2$ denote the image domain. A quantitative cardiac MRI baseline sequence consists of $L$ frames acquired under different settings, represented as a set $\mathcal{I}=\{I_1,\dots,I_L\}$ with $I_i:\Omega\to\mathbb{R}$. Groupwise registration estimates dense displacement fields $\{\mathbf{u}_i\}_{i=1}^L$ and transformations $\phi_i:\Omega\to\Omega$ defined by $\phi_i(x)=x+\mathbf{u}_i(x),\, x\in\Omega$, mapping all frames to a shared canonical coordinate system. Let $\Phi=\{\phi_1,\dots,\phi_L\}$ and $\hat I_i=I_i\circ\phi_i$ denote the transformation set and aligned images, respectively.

The alignment facilitates downstream voxel-wise quantitative mapping. Under ideal alignment, tissue properties $\psi(x)$ at location $x$ are estimated by
\begin{equation}
\hat\psi(x)=\arg\min_{\psi}\sum_{i=1}^L \ell\big(\hat I_i(x),\,\mathcal{M}_i(\psi(x))\big), \quad x\in\Omega,
\label{eq:qcmr_fitting_objective}
\end{equation}
where $\mathcal{M}_i$ is the signal model for the $i$-th frame under its acquisition settings (Fig.~\ref{data}), and $\ell$ is a pointwise loss. Eq.~\ref{eq:qcmr_fitting_objective} suggests two structural requirements for registration: it should generalize across varying $L$, and respect permutation symmetry over frame indices $\{1,\dots,L\}$.

\begin{figure}[t]
    \centering
    \includegraphics[width=\linewidth]{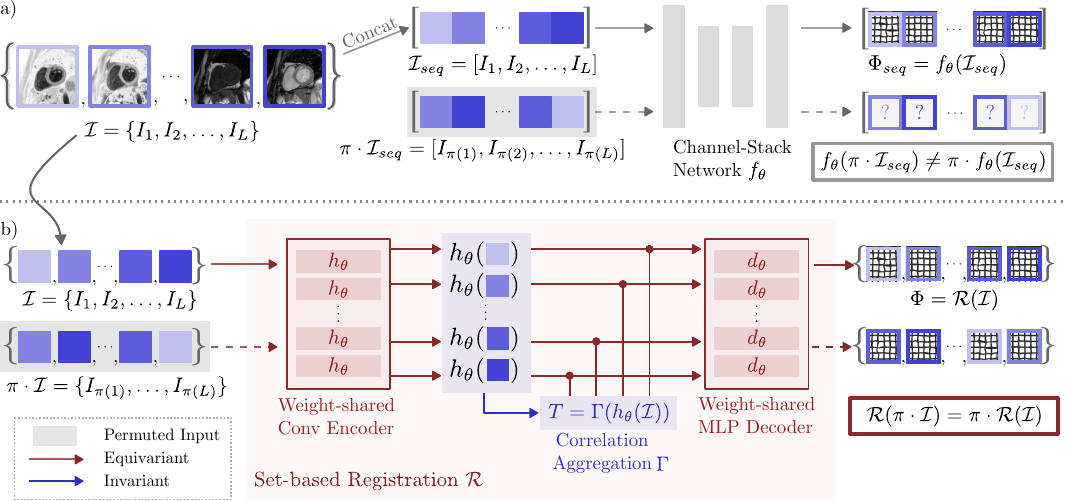}
    \caption{Comparison of two learning-based groupwise registration designs: (a) The conventional channel stacking design concatenates frames; it is order-sensitive ($f_\theta(\pi \cdot \mathcal{I}_{seq}) \neq \pi \cdot f_\theta(\mathcal{I}_{seq})$) and tied to a fixed length $L$. (b) Set-based design (\AnyTwoReg) uses shared encoders and a canonical reference $T=\Gamma(\mathcal{I})$ broadcasted to all frames; it enables permutation-equivariant registration ($\mathcal{R}(\pi \cdot \mathcal{I})=\pi \cdot \mathcal{R}(\mathcal{I})$) for an arbitrary $L$.}
    \label{fig:placeholder}
\end{figure}
\subsection{Permutation-equivariant registration network}
\label{sec:network_overview}
To handle varying $L$ and frame-index symmetry, the registration operator $\mathcal{R}: \mathcal{I} \mapsto \Phi$ should be permutation-equivariant:
\begin{equation}
    \mathcal{R}(\pi \cdot \mathcal{I}) = \pi \cdot \mathcal{R}(\mathcal{I}),\qquad \forall \pi \in S_L .
\end{equation}

That is, permuting the input frames should only permute the predicted transformations. In groupwise registration, this calls for an order-invariant canonical context to which all frames align. We enforce this with an invariant-to-equivariant design: a set operator $\Gamma$ constructs a reference, $\Gamma(\pi \cdot \mathcal{I})=\Gamma(\mathcal{I})$, which is broadcast to all framewise weight-shared branches, thus permutation-equivariant. 

Based on this principle, we propose \AnyTwoReg{} (Fig.~\ref{fig:placeholder}b), a set-based multi-scale registration network with $K$ scales ($k=1$ the finest). At scale $k$, a frame-shared encoder extracts per-frame features $\{z_i^{(k)}\}_{i=1}^L$. An invariant operator $\Gamma^{(k)}$ aggregates a canonical feature $T^{(k)}$ from $\{z_i^{(k)}\}$ and broadcasts it to all frames. The concatenated features $[z_i^{(k)},T^{(k)}]$ update the next-scale features and drive a shared CorrMLP-style decoder~\cite{meng2024correlation} to predict incremental transformations $\phi_i^{(k)}$. The final transformation composes predictions from coarse to fine:
\begin{equation}
\phi_i
=
\phi_i^{(1)}
\circ \uparrow\!\bigl(\phi_i^{(2)}\bigr)
\circ \cdots \circ
\uparrow\!\bigl(\phi_i^{(K)}\bigr),
\label{eq:phi_composition}
\end{equation}
where $\uparrow(\cdot)$ upsamples to the finest grid. Sec.~\ref{sec:set_aggregation} details our instantiation of $\Gamma^{(k)}$.
\subsection{Correlation-guided set aggregation}
\label{sec:set_aggregation}
We instantiate the scale-specific invariant operator $\Gamma^{(k)}$ using a correlation-guided principal component (PC) aggregation as in~\cite{huizinga2016pca,polfliet2018intrasubject}. 

Given scale-$k$ features $z_i^{(k)}\in\mathbb{R}^{C_k\times H_k\times W_k}$, we form vectorized $\ell_2$-normalized representations $\tilde z_i^{(k)}$ and compute the frame-wise correlation matrix $\mathbf{C}^{(k)}\in\mathbb{R}^{L\times L}$. Let $v^{(k)}$ be the leading eigenvector (first PC) of $\mathbf{C}^{(k)}$, which captures the principal mode of variation across the sequence. We then aggregate the canonical feature using the entries $v_i^{(k)}$ of the $v^{(k)}$:
\begin{equation}
T^{(k)} = \sum_{i=1}^L w_i^{(k)} z_i^{(k)},\qquad 
w_i^{(k)} = \frac{v_i^{(k)}}{\sum_{j=1}^L v_j^{(k)}}.
\label{eq:canonical_feature_corr}
\end{equation}
Under a frame-wise index permutation, $v_i^{(k)}$ (and thus $w_i^{(k)}$) permutes accordingly. Therefore, with a deterministic eigenvector sign convention of $\sum_{j=1}^L v_j^{(k)} > 0$ is permutation\allowbreak-invariant, leading to $\Gamma^{(k)}\!\left(\{z_i^{(k)}\}_{i=1}^L\right)$.

To inject this canonical context back to each frame, we broadcast $T^{(k)}$ and apply a shared update:
\begin{equation}
z_i^{(k+1)}=\sigma\!\left(\mathrm{Conv}^{(k)}\!\left([z_i^{(k)},\,T^{(k)}]\right)\right),
\label{eq:equivariant_update_corr}
\end{equation}
which preserves permutation equivariance since $T^{(k)}$ is shared across frames and $\mathrm{Conv}^{(k)}$ uses shared weights. The set aggregation operates on a relatively small $L\times L$ matrix, and therefore adds negligible overhead compared with per-frame convolutions. This is further empirically validated in Sec. \ref{sec:scalability}.

\subsection{Auxiliary anatomy-driven feature stream}
\label{sec:aux_fm}
Anatomy-driven, contrast-agnostic features have been shown to improve learning-based registration~\cite{voxelmorph,hanania2023pcmc}; when available from a pretrained model (e.g., a segmentation foundation model), they can be optionally incorporated as an auxiliary stream. Let $g(\cdot)$ be a frozen feature extractor. For each frame $I_i$, we extract
\begin{equation}
F_i = g(I_i), \quad F_i \in \mathbb{R}^{C_f \times H_f \times W_f}.
\end{equation}
Applying $g$ independently to each frame yields a feature set $\mathcal{F}=\{F_1,\dots,F_L\}$ that preserves the set structure and permutation equivariance.

The auxiliary stream is encoded in parallel by a frame-shared multi-scale encoder $h'^{(k)}_\theta$. At each scale, we fuse pretrained features with image-branch features via a lightweight convolution before applying $\Gamma^{(k)}$. This leads to set aggregation without altering the permutation-equivariant formulation.
\subsection{Loss functions}

Using the final warped images $\hat I_i = I_i \circ \phi_i$, a correlation-weighted template is constructed as
$
T_I = \sum_{i=1}^L w_i\,\hat I_i
$
where weights $\{w_i\}$ are computed via the correlation-guided weighting scheme (Sec.~\ref{sec:set_aggregation}). We use conditional template entropy (CTE)~\cite{polfliet2018intrasubject} for multi-contrast registration as the dissimilarity function:
\begin{equation}
\mathcal{L}_{\mathrm{CTE}}(\hat{\mathcal{I}},T_I)
= \frac{1}{L}\sum_{i=1}^L H\!\left(T_I \mid \hat I_i\right),
\qquad
H(T\mid I)=H(T,I)-H(I).
\end{equation}
When the auxiliary feature stream is utilized, the warped feature maps $\hat F_i = F_i\circ\phi_i$ yield the feature template $T_F = \sum_i w_{f,i} \hat F_i$. The weights $\{w_{f,i}\}$ are independently computed via correlation in the feature space. The total loss is:
\begin{equation}
\mathcal{L} = \mathcal{L}_{\mathrm{CTE}}(\hat{\mathcal{I}}, T_I) + \lambda_f \mathcal{L}_{\mathrm{CTE}}(\hat{\mathcal{F}}, T_F) + \lambda_s \mathcal{L}_{\text{smooth}}(\Phi) + \lambda_c \mathcal{L}_{\text{cyclic}}(\Phi),
\end{equation}
where $\mathcal{L}_{\text{smooth}}$ penalizes displacement gradients to ensure deformation regularity and $\mathcal{L}_{\text{cyclic}}$~\cite{zhang2021groupregnet} suppresses non-zero group motion to prevent spatial drift.

\section{Experiments}
\noindent\textbf{Data.} We evaluate four heterogeneous cardiac MRI datasets comprising 2D short-axis slices (Fig.~\ref{data}). In-domain training uses the public STONE dataset (free-breathing slice-interleaved $T_1$ mapping; 210 subjects; 1050 slices, 150/30/30 train/val/test subject-wise split, $L=11$)~\cite{weingartner2015free,el2018nonrigid}. For zero-shot evaluation, we use an in-house MOLLI dataset (modified Look-Locker $T_1$ mapping; 10 subjects, 60 slices, pre-/post-contrast, $L=11$)~\cite{messroghli2004modified} and an in-house ASL dataset (myocardial arterial spin labeling; 6 subjects; $L=60$ via 12 images$\times$5 repetitions)~\cite{bovzic2024improved}. Additionally, we include the public ACDC dataset (bSSFP Cine; test split; 25 subjects, 256 slices; $L\in[10,30]$)~\cite{bernard2018deep}, which is not a quantitative sequence and contains regular cardiac motion, for an out-of-distribution stress test.

\noindent\textbf{Evaluation metrics.} We report slice-wise all-pair Dice for left ventricle (LV) and myocardium (Myo) on STONE (expert annotations), and for LV, right ventricle (RV), and Myo on MOLLI, ASL, and Cine using ACDC-trained nnU-Net predictions~\cite{isensee2021nnu} independent of the feature stream. We manually annotate RV insertion points on STONE and MOLLI to measure target registration error (TRE). On STONE, we further assess $T_1$ mapping quality via per-pixel curve-fitting $R^2$ and deformation regularity via $\mathrm{std}(\log\det J)$.

\noindent\textbf{Baselines and ablations.} We compare with Elastix~\cite{klein2009elastix} (optimization-based; PCA2 similarity~\cite{huizinga2016pca}; B-spline; 4-pixel spacing), GroupRegNet~\cite{zhang2021groupregnet} (channel-stacked, one-shot), PCA-Relax~\cite{zhang2024deep} (channel-stacked learning with PCA2), and MultiMorph~\cite{abulnaga2025multimorph} adapted to groupwise registration, including its segmentation-guided variant (MultiMorph-Seg). We also evaluate \AnyTwoReg without foundation model features, replace correlation with mean aggregation for ablation. We include an instance-optimization (IO)~\cite{zhang2024deep} variant that refines test cases for 30 inference steps.

\noindent\textbf{Implementation.} Images are center-cropped to $192\times192$. The encoder utilizes $K=4$ scales with 16 channels. Auxiliary anatomical features are derived from a pretrained Reverse Imaging segmentation foundation model~\cite{zhao2025reverse} trained on the ACDC train split. For MultiMorph-Seg, official STONE annotations are used. Models are trained for 50 epochs via AdamW (lr $10^{-4}$) with $\lambda_f=0.5$, $\lambda_s=10$, and $\lambda_c=0.05$. For fair comparison, MultiMorph optimizes the same CTE loss; all hyperparameters are tuned on the STONE validation with comparable $\mathrm{std}(\log\det J)$ and folding ratios $<10^{-5}$. One-shot GroupRegNet optimizes from random initialization for 250 steps. \AnyTwoReg IO refines for 30 steps (lr $3\times10^{-5}$). Implementation used PyTorch 2.8.0 on an NVIDIA RTX5090. Code and weights are available at \url{https://github.com/YiZhang025/Any2Reg}.

\begin{table*}[!tb]
\centering
\caption{Main quantitative results for in-domain (STONE) and three cardiac MRI datasets. $\sigma_{\log|J|} \equiv \mathrm{std}(\log\det J)\times 10^{-2}$. \failrun~indicates not applicable due to sequence length constraints; \bestlegend/\secondlegend{} mark the best and second-best results. }
\label{tab:main_comparison}
\setlength{\tabcolsep}{1pt}
{\fontsize{8}{9.6}\selectfont
\begin{tabular}{@{}l cc c cc c c@{}}
\toprule
\multirow{2.5}{*}{\textbf{Method}} & \multicolumn{3}{c}{\textbf{STONE}} & \multicolumn{2}{c}{\textbf{MOLLI}} & \textbf{ASL} & \textbf{Cine} \\
\cmidrule(lr){2-4} \cmidrule(lr){5-6} \cmidrule(lr){7-7} \cmidrule(lr){8-8} 
 & Dice $\uparrow$ & TRE $\downarrow$ &  $\sigma_{\log|J|}$ $\downarrow$ & Dice $\uparrow$ & TRE $\downarrow$ & Dice $\uparrow$ & Dice $\uparrow$ \\
\midrule
Raw & \mstd{78.8}{7.1} & \mstd{4.19}{1.26} & / & \mstd{78.6}{14.8} & \mstd{3.16}{1.50} & \mstd{76.7}{7.8} & \mstd{79.1}{4.8} \\
Elastix & \mstd{85.0}{5.4} & \mstd{3.46}{1.08} & \mstd{12.7}{4.5} & \mstd{79.2}{14.3} & \mstd{3.10}{1.32} & \mstd{79.4}{8.1} & \mstd{86.8}{3.8} \\
GroupRegNet & \mstd{83.0}{6.2} & \mstd{3.60}{1.04} & \secondcell{\mstd{9.3}{3.5}} & \mstd{79.7}{14.6} & \mstd{3.09}{1.36} & \mstd{78.6}{8.5} & \mstd{80.5}{4.7} \\
PCA-Relax & \mstd{80.6}{6.6} & \mstd{4.01}{1.16} & \mstd{14.0}{4.7} & \mstd{78.2}{14.8} & \mstd{3.13}{1.49} & \failrun & \failrun \\
MultiMorph & \mstd{83.1}{6.4} & \mstd{3.77}{1.15} & \mstd{14.7}{1.9} & \mstd{78.4}{14.3} & \mstd{3.11}{1.39} & \mstd{75.2}{5.7} & \mstd{84.8}{4.3} \\
MultiMorph-Seg & \mstd{85.0}{5.9} & \mstd{3.71}{1.17} & \mstd{15.2}{2.2} & \mstd{78.8}{14.1} & \mstd{3.18}{1.36} & \mstd{76.9}{6.4} & \mstd{86.1}{3.9} \\
\midrule
\AnyTwoReg{} w/o FM & \mstd{84.2}{5.9} & \mstd{3.65}{1.11} & \bestcell{\mstd{8.3}{1.4}} & \mstd{78.9}{14.5} & \mstd{3.17}{1.43} & \mstd{79.0}{7.2} & \mstd{89.5}{2.7} \\
\AnyTwoReg{} mean & \mstd{86.4}{4.8} & \mstd{3.23}{0.86} & \mstd{14.2}{1.6} & \mstd{84.7}{14.7} & \mstd{3.02}{0.97} & \mstd{80.3}{7.7} & \secondcell{\mstd{89.9}{2.5}} \\
\AnyTwoReg & \secondcell{\mstd{86.9}{4.7}} & \secondcell{\mstd{3.18}{0.88}} & \mstd{12.3}{3.0} & \secondcell{\mstd{84.8}{14.8}} & \bestcell{\mstd{2.96}{0.96}} & \secondcell{\mstd{80.5}{7.9}} & \mstd{89.8}{2.5} \\
\AnyTwoReg{} IO & \bestcell{\mstd{86.9}{4.4}} & \bestcell{\mstd{3.14}{0.83}} & \mstd{13.7}{3.1} & \bestcell{\mstd{84.9}{15.0}} & \secondcell{\mstd{2.98}{0.95}} & \bestcell{\mstd{80.6}{7.6}} & \bestcell{\mstd{90.9}{2.1}} \\
\bottomrule
\end{tabular}%
}
\end{table*}

\begin{figure*}[b]
\centering
\includegraphics[width=\textwidth]{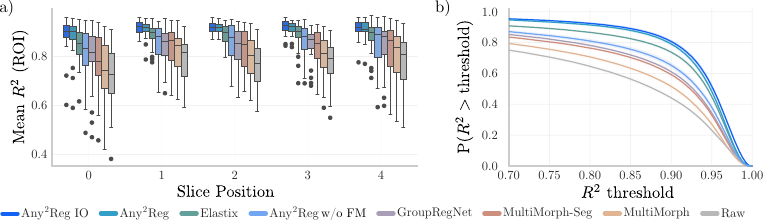}
\caption{$T_1$ mapping quality on STONE (LV+Myo). (a) Slice-wise mean $R^2$. Slice position: base (0) to apex (4). \AnyTwoReg{} IO demonstrates the best alignment at all slice positions. (b) $R^2$ survival curves show \AnyTwoReg{} has consistently higher fitting quality as a result of superior alignment.}
\label{fig:r2_survival}
\end{figure*}

\section{Results and Discussion}

\noindent{\textbf{Main performance.}} Table~\ref{tab:main_comparison} presents quantitative results for in-domain and zero-shot cross-protocol scenarios. On STONE, \AnyTwoReg{}+IO achieves the highest Dice and lowest TRE. The default \AnyTwoReg{} consistently outperforms U-Net baselines across all datasets. Ablation on aggregation shows \AnyTwoReg{} surpasses mean aggregation (\AnyTwoReg{} mean) in quantitative MRI datasets, validating that correlation yields a more informative canonical representation under contrast variation. \AnyTwoReg{} w/o FM also outperforms baselines in zero-shot cases; this confirms that set-based aggregation improves generalization independently of the anatomy-driven features. Strong zero-shot performance on Cine shows robust generalization to a completely different spatiotemporal sequence. In contrast, fixed-length channel-stacked method PCA-Relax fails on ASL and Cine due to sequence-length constraints. 

\noindent{\textbf{Scalability analysis.}} We evaluated \AnyTwoReg's scalability up to $L=512$. \AnyTwoReg{} inference scales linearly ($\sim$2.4 ms per additional image), taking $\sim$0.2 s at $L=64$ and $\sim$1.2 s at $L=512$. The correlation aggregation GPU runtime overhead remains negligible ($<4\%$). The 30-step IO variant also scales linearly ($\sim$12 s at $L=64$). Conversely, conventional method (Elastix) grows superlinearly ($\sim$70 s at $L=64$; $\sim$260 s at $L=128$).

\begin{figure*}[!t]
\centering
\includegraphics[width=\textwidth]{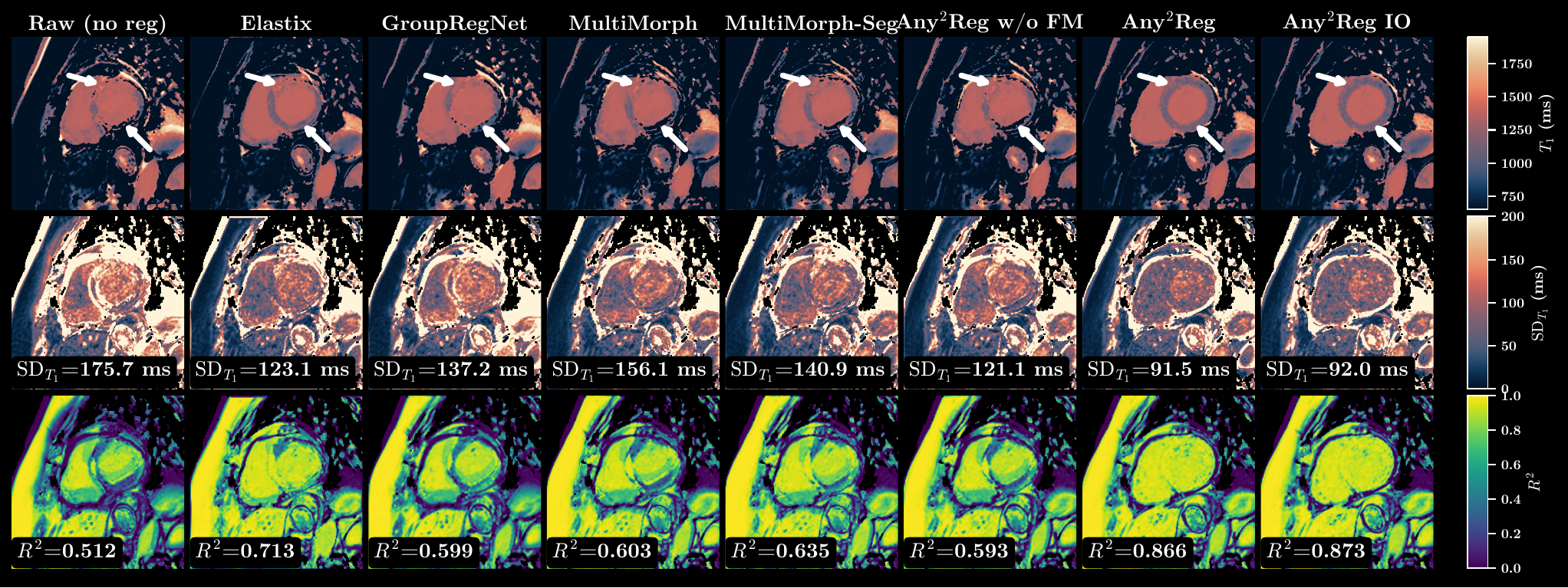}
\caption{Qualitative result on a representative STONE sequence with $T_1$ map, fitting uncertainty $\mathrm{SD}_{T_1}$ ($\downarrow$), and pixelwise fitting $R^2$ ($\uparrow$). White arrows indicate the location of large motion. Both \AnyTwoReg{} and \AnyTwoReg{} IO yield clear $T_1$ maps with high precision.}
\label{fig:head2head_0138}
\end{figure*}

\noindent{\textbf{Quantitative Mapping.}} Figure~\ref{fig:r2_survival} illustrates the benefits for downstream $T_1$ mapping, wherein \AnyTwoReg{} maintains the highest $R^2$ survival rates. The most substantial margin is observed in the base segment, where motion degrades alignment. Qualitatively, as shown in Figure~\ref{fig:head2head_0138}, imprecise registration mixes the signal contributions from adjacent tissues across frames. This degrades the quality of fit, resulting in reduced $R^2$ values and high $T_1$ fitting uncertainty ($\mathrm{SD}_{T_1}$)~\cite{kellman2013t1}. \AnyTwoReg{} effectively resolves these motion artifacts, achieving a superior fit ($R^2=0.866$, $\mathrm{SD}_{T_1}=91.5$\,ms) in high precision (low $\mathrm{SD}_{T_1}$) $T_1$ maps.

\label{sec:scalability}
\section{Conclusion}

We propose a new way of groupwise registration, \AnyTwoReg, which can operate on variable-length, variable-contrast image sequences of cardiac MRI. \AnyTwoReg reformulates groupwise registration as a set-to-set prediction problem, and completely eliminates the rigid coupling of sequence length and image order in conventional channel-stacked architectures. Trained solely on one quantitative MRI sequence, \AnyTwoReg achieves robust zero-shot generalization across heterogeneous MRI protocols, including unseen quantitative sequences and spatiotemporal Cine sequences. It improves spatial alignment and downstream quantitative mapping over state-of-the-art baselines while maintaining speed advantage. \AnyTwoReg{} provides a highly scalable and robust solution for registering cardiac MRI, and its generic setup enables extension to other medical image registration applications. 

\clearpage
\bibliographystyle{splncs04}
\bibliography{example}

@article{bovzic2024improved,
  title={Improved reproducibility for myocardial ASL: Impact of physiological and acquisition parameters},
  author={Bo{\v{z}}i{\'c}-Iven, Ma{\v{s}}a and Rapacchi, Stanislas and Tao, Qian and Pierce, Iain and Thornton, George and Nitsche, Christian and Treibel, Thomas A and Schad, Lothar R and Weing{\"a}rtner, Sebastian},
  journal={Magnetic Resonance in Medicine},
  volume={91},
  number={1},
  pages={118--132},
  year={2024},
  publisher={Wiley Online Library}
}

@inproceedings{zhao2025reverse,
  title={Reverse Imaging for Wide-spectrum Generalization of Cardiac MRI Segmentation},
  author={Zhao, Yidong and Zhang, Yi and Simonetti, Orlando and Han, Yuchi and Tao, Qian},
  booktitle={International Conference on Medical Image Computing and Computer-Assisted Intervention},
  year={2025},
  organization={Springer}
}

@article{martin2020groupwise,
  title={Groupwise non-rigid registration with deep learning: an affordable solution applied to 2D cardiac cine MRI reconstruction},
  author={Mart{\'\i}n-Gonz{\'a}lez, Elena and Sevilla, Teresa and Revilla-Orodea, Ana and Casaseca-de-la-Higuera, Pablo and Alberola-L{\'o}pez, Carlos},
  journal={Entropy},
  volume={22},
  number={6},
  pages={687},
  year={2020},
  publisher={MDPI}
}

@article{klein2009elastix,
  title={Elastix: a toolbox for intensity-based medical image registration},
  author={Klein, Stefan and Staring, Marius and Murphy, Keelin and Viergever, Max A and Pluim, Josien PW},
  journal={IEEE transactions on medical imaging},
  volume={29},
  number={1},
  pages={196--205},
  year={2009},
  publisher={IEEE}
}

@inproceedings{arava2021deep,
  title={Deep-Learning based Motion Correction for Myocardial T1 Mapping},
  author={Arava, Dar and Masarwy, Mohammad and Khawaled, Samah and Freiman, Moti},
  booktitle={IEEE International Conference on Microwaves, Antennas, Communications and Electronic Systems},
  pages={55--59},
  year={2021}
}

@article{li2022motion,
  title={Motion correction for native myocardial T1 mapping using self-supervised deep learning registration with contrast separation},
  author={Li, Yuze and Wu, Chunyan and Qi, Haikun and Si, Dongyue and Ding, Haiyan and Chen, Huijun},
  journal={NMR in Biomedicine},
  volume={35},
  number={10},
  pages={e4775},
  year={2022},
  publisher={Wiley Online Library}
}

@article{haaf2016cardiac,
  title={Cardiac T1 mapping and extracellular volume (ECV) in clinical practice: a comprehensive review},
  author={Haaf, Philip and Garg, Pankaj and Messroghli, Daniel R and Broadbent, David A and Greenwood, John P and Plein, Sven},
  journal={Journal of Cardiovascular Magnetic Resonance},
  volume={18},
  number={1},
  pages={89},
  year={2016},
  publisher={Elsevier}
}

@article{zhang2021groupregnet,
  title={GroupRegNet: a groupwise one-shot deep learning-based 4D image registration method},
  author={Zhang, Yunlu and Wu, Xue and Gach, H Michael and Li, Harold and Yang, Deshan},
  journal={Physics in Medicine \& Biology},
  volume={66},
  number={4},
  pages={045030},
  year={2021},
  publisher={IOP Publishing}
}

@inproceedings{hanania2023pcmc,
  title={PCMC-T1: Free-Breathing Myocardial T1 Mapping with Physically-Constrained Motion Correction},
  author={Hanania, Eyal and Volovik, Ilya and Barkat, Lilach and Cohen, Israel and Freiman, Moti},
  booktitle={International Conference on Medical Image Computing and Computer-Assisted Intervention},
  pages={226--235},
  year={2023},
  organization={Springer}
}

@article{voxelmorph,
  title={VoxelMorph: a learning framework for deformable medical image registration},
  author={Balakrishnan, Guha and Zhao, Amy and Sabuncu, Mert R and Guttag, John and Dalca, Adrian V},
  journal={IEEE transactions on medical imaging},
  volume={38},
  number={8},
  pages={1788--1800},
  year={2019},
  publisher={IEEE}
}

@article{tao2018robust,
  title={Robust motion correction for myocardial T1 and extracellular volume mapping by principle component analysis-based groupwise image registration},
  author={Tao, Qian and van der Tol, Pieternel and Berendsen, Floris F and Paiman, Elisabeth HM and Lamb, Hildo J and van der Geest, Rob J},
  journal={Journal of Magnetic Resonance Imaging},
  volume={47},
  number={5},
  pages={1397--1405},
  year={2018},
  publisher={Wiley Online Library}
}

@article{makela2002review,
  title={A review of cardiac image registration methods},
  author={Makela, Timo and Clarysse, Patrick and Sipila, Outi and Pauna, Nicoleta and Pham, Quoc Cuong and Katila, Toivo and Magnin, Isabelle E},
  journal={IEEE Transactions on medical imaging},
  volume={21},
  number={9},
  pages={1011--1021},
  year={2002},
  publisher={IEEE}
}

@article{kellman2014t1,
  title={T1-mapping in the heart: accuracy and precision},
  author={Kellman, Peter and Hansen, Michael S},
  journal={Journal of cardiovascular magnetic resonance},
  volume={16},
  pages={1--20},
  year={2014},
  publisher={Springer}
}

@article{xue2012motion,
  title={Motion correction for myocardial T1 mapping using image registration with synthetic image estimation},
  author={Xue, Hui and Shah, Saurabh and Greiser, Andreas and Guetter, Christoph and Littmann, Arne and Jolly, Marie-Pierre and Arai, Andrew E and Zuehlsdorff, Sven and Guehring, Jens and Kellman, Peter},
  journal={Magnetic resonance in medicine},
  volume={67},
  number={6},
  pages={1644--1655},
  year={2012},
  publisher={Wiley Online Library}
}

@inproceedings{li2023contrast,
  title={Contrast-Agnostic Groupwise Registration by Robust PCA for Quantitative Cardiac MRI},
  author={Li, Xinqi and Zhang, Yi and Zhao, Yidong and van Gemert, Jan and Tao, Qian},
  booktitle={International Workshop on Statistical Atlases and Computational Models of the Heart},
  pages={77--87},
  year={2023}
}

@article{kellman2013t1,
  title={T1 and extracellular volume mapping in the heart: estimation of error maps and the influence of noise on precision},
  author={Kellman, Peter and Arai, Andrew E and Xue, Hui},
  journal={Journal of Cardiovascular Magnetic Resonance},
  volume={15},
  number={1},
  pages={1--12},
  year={2013},
  publisher={BioMed Central}
}

@inproceedings{zhang2024deep,
  title={Deep-Learning-Based Groupwise Registration for Motion Correction of Cardiac T 1 Mapping},
  author={Zhang, Yi and Zhao, Yidong and Huang, Lu and Xia, Liming and Tao, Qian},
  booktitle={International Conference on Medical Image Computing and Computer-Assisted Intervention},
  pages={586--596},
  year={2024},
  organization={Springer}
}

@article{huizinga2016pca,
  title={PCA-based groupwise image registration for quantitative MRI},
  author={Huizinga, Wyke and Poot, Dirk HJ and Guyader, J-M and Klaassen, Remy and Coolen, Bram F and van Kranenburg, Matthijs and Van Geuns, RJM and Uitterdijk, Andr{\'e} and Polfliet, Mathias and Vandemeulebroucke, Jef and others},
  journal={Medical image analysis},
  volume={29},
  pages={65--78},
  year={2016},
  publisher={Elsevier}
}

@article{messroghli2004modified,
  title={Modified Look-Locker inversion recovery (MOLLI) for high-resolution T1 mapping of the heart},
  author={Messroghli, Daniel R and Radjenovic, Aleksandra and Kozerke, Sebastian and Higgins, David M and Sivananthan, Mohan U and Ridgway, John P},
  journal={Magnetic Resonance in Medicine},
  volume={52},
  number={1},
  pages={141--146},
  year={2004},
  publisher={Wiley Online Library}
}

@inproceedings{ronneberger2015u,
  title={U-net: Convolutional networks for biomedical image segmentation},
  author={Ronneberger, Olaf and Fischer, Philipp and Brox, Thomas},
  booktitle={Medical Image Computing and Computer-Assisted Intervention, October 5-9, 2015, Part III 18},
  pages={234--241},
  year={2015},
  organization={Springer}
}

@article{campello2021multi,
  title={Multi-centre, multi-vendor and multi-disease cardiac segmentation: the M\&Ms challenge},
  author={Campello, Victor M and Gkontra, Polyxeni and Izquierdo, Cristian and Martin-Isla, Carlos and Sojoudi, Alireza and Full, Peter M and Maier-Hein, Klaus and Zhang, Yao and He, Zhiqiang and Ma, Jun and others},
  journal={IEEE Transactions on Medical Imaging},
  volume={40},
  number={12},
  pages={3543--3554},
  year={2021},
  publisher={IEEE}
}

@article{hanania2025mbss,
  title={MBSS-T1: Model-based subject-specific self-supervised motion correction for robust cardiac T1 mapping},
  author={Hanania, Eyal and Zehavi-Lenz, Adi and Volovik, Ilya and Link-Sourani, Daphna and Cohen, Israel and Freiman, Moti},
  journal={Medical Image Analysis},
  volume={102},
  pages={103495},
  year={2025},
  publisher={Elsevier}
}

@article{bernard2018deep,
  title={Deep learning techniques for automatic MRI cardiac multi-structures segmentation and diagnosis: is the problem solved?},
  author={Bernard, Olivier and Lalande, Alain and Zotti, Clement and Cervenansky, Frederick and Yang, Xin and Heng, Pheng-Ann and Cetin, Irem and Lekadir, Karim and Camara, Oscar and Ballester, Miguel Angel Gonzalez and others},
  journal={IEEE transactions on medical imaging},
  volume={37},
  number={11},
  pages={2514--2525},
  year={2018},
  publisher={ieee}
}

@article{he2025instantgroup,
  title={InstantGroup: Instant Template Generation for Scalable Group of Brain MRI Registration},
  author={He, Ziyi and Chung, Albert CS},
  journal={IEEE Transactions on Image Processing},
  year={2025},
  publisher={IEEE}
}

@inproceedings{abulnaga2025multimorph,
  title={Multimorph: On-demand atlas construction},
  author={Abulnaga, S Mazdak and Hoopes, Andrew and Dey, Neel and Hoffmann, Malte and Fischl, Bruce and Guttag, John and Dalca, Adrian},
  booktitle={Proceedings of the Computer Vision and Pattern Recognition Conference},
  pages={30906--30917},
  year={2025}
}

@article{weingartner2015free,
  title={Free-breathing multislice native myocardial T1 mapping using the slice-interleaved T1 (STONE) sequence},
  author={Weing{\"a}rtner, Sebastian and Roujol, S{\'e}bastien and Ak{\c{c}}akaya, Mehmet and Basha, Tamer A and Nezafat, Reza},
  journal={Magnetic resonance in medicine},
  volume={74},
  number={1},
  pages={115--124},
  year={2015},
  publisher={Wiley Online Library}
}

@article{isensee2021nnu,
  title={nnU-Net: a self-configuring method for deep learning-based biomedical image segmentation},
  author={Isensee, Fabian and Jaeger, Paul F and Kohl, Simon AA and Petersen, Jens and Maier-Hein, Klaus H},
  journal={Nature methods},
  volume={18},
  number={2},
  pages={203--211},
  year={2021},
  publisher={Nature Publishing Group US New York}
}

@article{el2018nonrigid,
  title={Nonrigid active shape model--based registration framework for motion correction of cardiac T1 mapping},
  author={El-Rewaidy, Hossam and Nezafat, Maryam and Jang, Jihye and Nakamori, Shiro and Fahmy, Ahmed S and Nezafat, Reza},
  journal={Magnetic resonance in medicine},
  volume={80},
  number={2},
  pages={780--791},
  year={2018},
  publisher={Wiley Online Library}
}

@article{polfliet2018intrasubject,
  title={Intrasubject multimodal groupwise registration with the conditional template entropy},
  author={Polfliet, Mathias and Klein, Stefan and Huizinga, Wyke and Paulides, Margarethus M and Niessen, Wiro J and Vandemeulebroucke, Jef},
  journal={Medical image analysis},
  volume={46},
  pages={15--25},
  year={2018},
  publisher={Elsevier}
}

@inproceedings{meng2024correlation,
  title={Correlation-aware coarse-to-fine mlps for deformable medical image registration},
  author={Meng, Mingyuan and Feng, Dagan and Bi, Lei and Kim, Jinman},
  booktitle={Proceedings of the IEEE/CVF Conference on Computer Vision and Pattern Recognition},
  pages={9645--9654},
  year={2024}
}

\end{document}